\documentstyle[aps]{revtex}
\begin{document}
\draft

\wideabs{
\title{
Numerical Relativity for Inspiraling Binaries in Co-Rotating Coordinates:
\\ Test Bed for Lapse and Shift Equations 
}

\author{Kip S.\ Thorne} 

\address{
Theoretical Astrophysics, California Institute of Technology,
Pasadena, CA 91125, and}
\address{
Max-Planck-Institut f\"ur GravitationsPhysik, Schlatzweg 1,
14473 Potsdam, Germany}
\date{Received 8 August 1998}
\maketitle
\begin{abstract}
Gravitational-wave data analysis requires a detailed understanding of
the highly relativistic, late stages of inspiral of neutron-star and black-hole
binaries.  A promising method to compute the late inspiral and its
emitted waves is numerical relativity in co-rotating coordinates.  The 
coordinates must be kept co-rotating via an appropriate choice of numerical
relativity's lapse and shift functions.  This article proposes a model 
problem
for testing the ability of various lapse and shift prescriptions to keep
the coordinates co-rotating.
\end{abstract}
}

\narrowtext

\section{Introduction}

Patrick Brady, Jolien Creighton and I \cite{ibbh} have discussed 
the importance, for gravitational-wave data analysis, of
understanding the evolution of inspiraling black-hole 
binaries.  We showed that there is a crucial gap between the point, during
inspiral, when a post-Newtonian analysis breaks down, and 
the point, just
before merger, when conventional numerical relativity techniques 
can take over the analysis.  We argued that this IBBH (Intermediate 
Binary Black Hole) gap might best be filled by numerical relativity
computations carried out using co-rotating coordinates.  In such coordinates,
the quantities being computed (components of the 3-metric $\gamma_{ij}$
and extrinsic curvature $K_{ij}$)
change on the timescale of inspiral $\tau_*$, which is much
longer than the orbital timescale $1/\Omega = 1/($orbital angular velocity).
Therefore, in such coordinates 
one may be able to take long time steps, constrained only
by $\tau_*$, thereby speeding up the computer
computations by a large factor relative to those carried out in the usual
asymptotically inertial coordinates.  

Numerical relativity in co-rotating coordinates may also be effective,
and perhaps necessary,
for analyses of the late stages of inspiral of neutron-star binaries and 
neutron-star / black-hole binaries.

A key foundation for such computations is a prescription for choosing the 
lapse and shift functions $\alpha$ and $\beta^i$, which generate the coordinate
system as the computation proceeds.  The shift function $\beta^i$ must keep the
coordinates co-rotating with the binary's orbital motion, and the lapse
function $\alpha$ must appropriately retard the time slicing in regions of
strengthening gravitational redshift; and this must be done quantitatively
in such a way that the combined effects of $\alpha$ and $\beta^i$ keep
$\partial \gamma_{ij}/\partial t \sim \gamma_{ij}/\tau_*$ and
$\partial K_{ij}/\partial t \sim K_{ij}/\tau_*$.  For simplicity, I 
refer to such coordinates as {\it co-rotating}, even though they require more
than just co-rotation. 
 
Brady, Creighton and I \cite{ibbh} have proposed a class of equations for the 
lapse and shift, any one of which (we argued) may keep the coordinates
co-rotating, if they have been chosen co-rotating to begin with.  The simplest
of these are the ``minimal-strain'' lapse and shift equations:
\begin{mathletters}
\label{eq:MinimalStrain}
\begin{equation}
D^j(D_{(i}\beta_{j)} - \alpha K_{ij}) = 0\;,
\label{eq:MinimalStrainBeta}
\end{equation}
in which $\alpha$ is to be replaced by the following expression in terms of
$\beta_j$
\begin{equation}
\alpha = {K^{ij}D_i\beta_j\over K^{ab}K_{ab}}\;.
\label{eq:MinimalStrainAlpha}
\end{equation}
\end{mathletters}
Here $D_i$ is the covariant derivative in the slice of constant time $t$, which
has 3-metric $\gamma_{ij}$ and extrinsic curvature $K_{ij}$.

Numerical relativists, in conversations with me, have suggested other choices
of lapse and shift that might keep the coordinates co-rotating.  A detailed
tests of their proposals and of Brady, Creighton's and mine are now 
needed.  The purpose of this
Brief Report is to suggest a model problem that can serve as a test bed for the
various proposed lapse and shift equations.  

This test bed is a specific 4-dimensional spacetime that 
resembles that of a fully relativistic, inspiraling compact binary.
More specifically, it is a spacetime with a 4-metric that
possesses, semi-quantitatively correctly, all those features
that are likely to significantly influence the behavior of the
proposed lapse and shift equations---with one exception:  
the test-bed 4-metric does not possess black-hole horizons.
Instead, it models a binary neutron-star system.  This is because horizons
may seriously complicate the task of implementing the proposed lapse and
shift equations; so in initial tests it is best to omit them.  It should
not be difficult to generalize my test-bed 4-metric to include horizons,
both nonrotating and rotating.  

Any lapse and shift equations
that stably generate and maintain a co-rotating coordinate system in my
test-bed spacetime will very probably do so also in a real 
numerical-relativity calculation---i.e., in a calculation that 
is simultaneously evolving the binary's true
spacetime geometry along with the coordinate system. 

\section{The Test-Bed Spacetime}

I shall describe the test-bed spacetime in a co-rotating coordinate system 
$\{T,X,Y,Z\}$ that must not be confused with the coordinates $\{t,x,y,z\}$ 
generated by some chosen lapse and shift equations.  Thinking of the spatial 
coordinates $\{X,Y,Z\}$ as Cartesian in some flat background metric, we can
introduce the flat-space
distance $\varpi$ from the binary's rotation axis, spherical polar 
coordinates $(R,\theta,\phi)$, and unit vectors $\hat\theta$ and $\hat\phi$
along the $\theta$ and $\phi$ directions:
\begin{eqnarray}
&&\varpi=\sqrt{X^2+Y^2}\;,\quad R = \sqrt{X^2+Y^2+Z^2}\;, \nonumber\\
&&\theta = \arcsin(\varpi/R)\;, \quad \phi = \arctan(Y/X)\;; \nonumber\\
&&\hat\theta = {\partial_\theta \over R}= {-\varpi \partial_Z + Z
\partial_\varpi\over R}\quad\hbox{where } \partial_\varpi = {X\partial_X
+ Y\partial_Y \over \varpi}\;, \nonumber\\
&&\hat\phi = {\partial_\phi\over\varpi} = 
{X\partial_Y - Y\partial_X \over\varpi}\;.
\label{eq:CoordinatesBases}
\end{eqnarray} 
Here $\{\partial_X, \partial_Y, \partial_Z\}$ are the usual Cartesian
coordinate basis vectors, $\{\partial_\theta, \partial_\phi\}$ are the usual
spherical coordinate basis vectors, and $\partial_\varpi$ is the unit vector
pointing away from the rotation axis in the flat background metric. 

The test-bed spacetime describes two identical stars in a circular orbit around
each other.  The spacetime's 4-metric depends on four parameters: the
time-independent mass $m$ of each star as measured gravitationally far from
the binary, the time-independent stellar radius $b$ and slowly evolving
stellar separation $2a$ as measured in the flat background metric, and 
the slowly evolving
orbital angular velocity $\Omega$ as measured far from the binary.  

We require that
\begin{equation} 
2m \alt b \alt a\;, \quad \Omega a \alt 1\;,
\label{eq:Inequalities}
\end{equation}
so the stars are larger than their Schwarzschild radii, their separation
is larger than the sum of their radii, and their orbital speeds are less
than the speed of light.  The ``approximately less than'' rather than strictly
less than is because $a$ and $b$ are parameters measured in the flat background
metric, not in the physical metric.  We also require that
\begin{equation}
{\partial a \over \partial T} \sim {a\over\tau_*}\;, \quad 
{\partial \Omega \over \partial T}
\sim {\Omega \over \tau_*}\;, \quad
\hbox{where } {\Omega\tau_*} \gg 1\;,
\label{eq:Timescales}
\end{equation}
so the inspiral is slow.  To be somewhat realistic, one might want to set
\begin{equation}
\Omega = \sqrt{2m\over(2a)^3}
\label{eq:AngularVelocity}
\end{equation}
(Kepler's law)
and let the separation $2a$ evolve in the manner dictated by the quadrupolar
description of radiation reaction \cite{mtw}
\begin{equation}
a(T) = a_o \left(1-{4T\over \tau_{*o}}\right)^{1/4}\;, 
\quad \hbox{where } \tau_{*o} = {5\over8}{a_o^4\over m^3}\;;
\label{eq:a(T)} 
\end{equation}
here $2a_o$ is the separation at time $T=0$, when the remaining
time to merger is $\Delta T = \tau_{*o}/4$, and the inspiral timescale is
$\tau_{*o} = a_o (d a_o / dT)^{-1}$.

The test-bed 4-metric is not actually a solution of Einstein's field equations,
but rather is cooked up to resemble the solutions that we expect numerical
calculations to reveal.  

We can write the 4-metric in 3+1 form in the co-rotating coordinates:
\begin{equation}
ds^2 = -A^2 dT^2 + \Gamma_{ij} (dX^i + B^i dT)(dX^j + B^j dT)\;.
\label{eq:4Metric}
\end{equation}
Here the co-rotating lapse $A$, shift $B^i$ and 3-metric 
$\Gamma_{ij}$ are not
to be confused with the lapse $\alpha$ and shift $\beta$ that are generated by
some proposed prescription for keeping the coordinates (nearly) co-rotating,
and the 3-metric $\gamma_{ij}$ of that prescription.  

The co-rotating $A$,
$B^i$ and $\Gamma_{ij}$ will depend explicitly on the spatial coordinates
$\{X,Y,Z\}$ and will depend on time $T$ solely through the slowly changing
separation $2a$ and angular velocity $\Omega$.  

To take account of the fact that information about the inspiral cannot
propagate faster than light, in the 4-metric we shall regard
$a$ and $\Omega$ 
as functions of (approximately) {\it retarded time} $T-R$ rather than time 
$T$: 
\begin{equation}
a = a(T-R)\;, \quad \Omega = \Omega(T-R)\;.
\label{eq:aOmega}
\end{equation}  
In numerical tests this might not be very important, since the
the full coordinate grid is not likely to be larger than a few
gravity-wave
wavelengths, $\sim 10/\Omega$, and thus will likely be much smaller
than $\tau_*$ (the timescale for changes of $a$ and $\Omega$), 
except possibly near
the end of a test.

We choose the (truly) co-rotating lapse $A$ to be $1$ minus the (approximate)
Newtonian potential of the binary, in accord with the Newtonian limit of
general relativity:
\begin{equation}
A = 1- m F_+\;,
\label{eq:A}
\end{equation}
where the function $F_+$ (``Newtonian potential per unit mass'') and an
associated function $F_-$ are defined by
\begin{eqnarray}
F_\pm &=& {1\over \left[ (X-a)^2 + Y^2 + Z^2 + b^2 \right]^{1/2}} 
\nonumber\\
&&\pm 
{1\over \left[ (X+a)^2 + Y^2 + Z^2 + b^2 \right]^{1/2}}\;,
\label{eq:F}
\end{eqnarray}
and $a \equiv a(T-R)$.  The $b^2$ rounds off the growth of the potential 
$F_\pm$ as one moves inside each star.

If our coordinates were inertial rather than co-rotating, the shift $\bf B$ 
would consist solely of a piece ${\bf B}_{\rm drag}$ produced by the dragging 
of inertial frames due to the binary's orbital motion.
The transformation $\bar\phi = \phi +\int\Omega dT$, 
from the inertial angular 
coordinate $\bar\phi$ to the co-rotating angular coordinate 
$\phi$, augments this frame-dragging shift 
by $\Omega\partial_\phi = \Omega\varpi\hat\phi$, thereby giving rise to our
following choice for the shift in co-rotating coordinates: 
\begin{equation}
{\bf B} = \left( 1- {4ma^2\over (R^2+a^2)^{3/2}} \right) \Omega \varpi
\hat \phi - {16ma^5\Omega \over(R^2+a^2)^2} F_- \partial_y\;. 
\label{eq:B}
\end{equation}

Our choice here of ${\bf B}_{\rm drag}$ consists of two pieces.  The first
piece, $-4ma^2 (R^2+a^2)^{-3/2} \Omega \varpi \hat \phi$, dominates at large
radii $R\gg a$, where it has the standard form and value 
$-(2J/R^3) \partial_\phi$
for the dragging of inertial frames by the binary's total angular momentum 
$J=2ma^2\Omega$.  
The second piece, $-16ma^5\Omega(R^2+a^2)^{-2}F_-\partial_y$
dominates in the vicinity of each star, where it has the standard form 
and value $-4{\bf p}/($distance to star) for the 
frame-dragging field produced by the star's linear momentum 
${\bf p} = \pm m a \Omega \partial_y$ 
($+$ for the star at $x=+a$; $-$ for the star at $-a$).
Because the vector fields $\partial_y$ and
$\varpi\hat\phi = \partial_\phi = X\partial_Y - Y\partial_X$ 
are both regular on the
rotation axis and at the origin, our chosen shift ${\bf B}$ is
regular.

We split our co-rotating 3-metric into two parts:  
\begin{equation}
\Gamma_{ij} = \Gamma^{\rm C}_{ij} + \Gamma^{\rm TT}_{ij}\;.
\label{eq:Gamma}
\end{equation}
Here
\begin{equation}
\Gamma^{\rm C}_{ij} = \delta_{ij} (1+m F_+)^2 
\label{eq:GammaC}
\end{equation}
is a conformally flat part
with the form (flat metric$)\times($1 - Newtonian potential)$^2$ 
expected from the
Newtonian limit of general relativity; 
\begin{eqnarray}
{\bf \Gamma}^{\rm TT} &=& {-4ma^2\Omega^3 (\Omega R)^4 \over (1+\Omega^2
R^2)^{5/2}} \nonumber\\
&&\times\left[ (1+\cos^2\theta) \cos(2\phi + 2\Omega R) 
(\hat\theta\hat\theta - \hat\phi \hat\phi) \right. 
\nonumber\\ 
&&
\left.\quad + 2 \cos\theta \sin(2\phi+ 2\Omega R)
(\hat\theta \hat\phi + \hat\phi \hat\theta) \right]
\label{eq:GammaTT}
\end{eqnarray}
is a transverse-traceless (TT)
part that represents outgoing gravitational waves in the radiation zone;
and two vectors written side by side, e.g. $\hat\theta \hat\phi$,
represent a tensor product.

The TT part of the 3-metric, ${\bf \Gamma}^{\rm TT}$,
requires discussion:  In the radiation zone, $R \gg 1/\Omega$, 
after transformation from the co-rotating angular coordinate $\phi$ to the
inertial-frame angular coordinate $\bar\phi = \phi + \int \Omega dT \simeq \phi
+ \Omega T$, ${\bf \Gamma}^{\rm TT}$
takes the standard quadrupole-moment form for outgoing gravitational waves 
produced by a nearly Newtonian binary:  ${\bf \Gamma}^{\rm TT} = h_+
(\hat\theta \hat\theta - \hat\phi \hat\phi) 
+ h_\times (\hat\theta
\hat\phi + \hat\phi \hat\theta)$. Here the waveforms $h_+$ and
$h_\times$ are \cite{apostolatos}
\begin{eqnarray}
h_+ &=& {-4ma^2\Omega^2\over R}(1+\cos^2\theta) \cos[2\bar\phi-2\Omega(T-R)]\;,
\nonumber\\
h_\times &=& {-4ma^2\Omega^2\over R}2\cos\theta \sin[2\bar\phi-2\Omega(T-R)]\;. 
\label{eq:WaveForms}
\end{eqnarray}
In co-rotating coordinates $\phi = \bar \phi -\Omega T$,
the time oscillation of this radiation field is
gone.  Instead of oscillating in time, it consists of a simple spiral pattern
in space that changes on the slow inspiral timescale $\tau_*$.  The fact that
it represents outgoing waves rather than ingoing shows up in 
the direction of the spiral and equivalently in the relation 
$(\partial_R - \Omega\partial_\phi)h_+ \sim
h_+/\tau_* \ll \Omega h_+$ and similarly for $h_\times$.  An ingoing-wave
pattern would exhibit the opposite direction of spiral and would have
$(\partial_R + \Omega\partial_\phi)h_+ \sim h_+/\tau_*$.

The specific functional form of ${\bf \Gamma}^{\rm TT}$, Eq.\
(\ref{eq:GammaTT}), is designed to have several properties: (i) It produces the
desired radiation field (\ref{eq:WaveForms})
at $R \gg 1/\Omega$. (ii) It becomes very small in the
near zone, $R\ll 1/\Omega$---negligibly small compared to the conformally flat
part of the 3-metric ${\bf \Gamma}^{\rm C}$.  (iii) It is regular on the
rotation axis, because as one approaches the 
axis $\cos(2\phi - 2\phi_o)(\hat\theta 
\hat\theta - \hat\phi \hat\phi) \pm \sin(2\phi - 2\phi_o)
(\hat\theta\hat\phi + \hat\phi\hat\theta)$ becomes a simple tidal
stretch along the $\phi_o$ direction.  (iv) It is regular at the origin, 
because its angular dependence is that of the electric-type tensor spherical
harmonic of order $l=2$, ${\bf T}^{E2,lm}$,
which is constructed via Clebsch-Gordon coupling
from scalar harmonics of order $l'=0,2$ and 4 \cite{rmp}, 
and the $R^4$ dependence at the
origin, multiplied by these scalar harmonics, produces regularity. 

There is no attempt, in the chosen expression for ${\bf \Gamma}^{\rm TT}$,
to represent, even approximately correctly, the radiation-reaction field inside
the binary.  This is because
the radiation-reaction field is so small inside the binary, 
$\sim \Omega^5 m a^2 R^2 \ll m/R$
\cite{mtwReact}, that
it is unlikely to have any significant influence on the behaviors of 
proposed lapses and shifts.  Similarly, there is no attempt to approximate
the details of the transition from near-zone field to radiation field. 
Except near the endpoint of inspiral, that transition field is small compared
to $m/R$, the important part of ${\bf \Gamma}^{\rm C}$.  

In summary, our test-bed 4-metric (\ref{eq:4Metric}) has the co-rotating
lapse (\ref{eq:A}), shift (\ref{eq:B}), and 3-metric 
(\ref{eq:Gamma})---(\ref{eq:GammaTT}). 

\section{Testing Proposed Lapses and Shifts}
\label{sec:Testing}

To test a proposed prescription for choosing the lapse $\alpha$ and shift
$\beta^i$---for example, the Brady-Creighton-Thorne minimal-strain prescription
(Eqs.\ (\ref{eq:MinimalStrain}) and associated boundary conditions discussed in
Ref.\ \cite{ibbh})---one can proceed as follows:

Choose the 3-surface $T=0$ as an initial 3-slice in spacetime, and label it
with the time coordinate $t=0$.  On this initial slice, set the
spatial coordinates $\{x,y,z\}$ equal to the co-rotating coordinates
$\{X,Y,Z\}$.  Correspondingly, the initial 3-metric and extrinsic curvature
in the $\{x,y,z\}$ coordinates will be
\begin{mathletters}
\begin{equation}
\gamma_{ij}(t=0) = \Gamma_{ij}(T=0)\;,
\label{eq:gammaInitial}
\end{equation}
\begin{equation} K_{ij} (t=0) = \left[ {1\over2A}
\left(-{\partial\Gamma_{ij}\over
\partial T} + 2 D_{(i} B_{j)}\right)\right]_{T=0}\;,
\label{eq:KInitial}
\end{equation}
\end{mathletters}
where the $T$-derivative acts solely through the dependence of $\Gamma_{ij}$ on
$a(T-R)$ and $\Omega(T-R)$.

Choose the initial lapse $\alpha(t=0)$ and shift $\beta^i(t=0)$ to be as close
as the proposed lapse and shift prescriptions allow to
\begin{equation}
\alpha(t=0) = A(T=0)\;, \quad \beta^i = B^i (T=0)\;.
\label{eq:alphabetaInitial}
\end{equation}
If the prescription does not allow these choices up to a fractional difference
of order $1/(\Omega\tau_*) \ll 1$, then the prescription is suspect.  
The minimal-strain prescription (\ref{eq:MinimalStrain}) does have this
property, as do all the other prescriptions discussed by Brady, Creighton and
Thorne \cite{ibbh}.

Beginning with these initial data, use the proposed prescription to
evolve forward in time
the lapse $\alpha$ and shift $\beta^i$, and the associated time slices
$t(T,X,Y,Z) = \hbox{constant}$ 
and the spatial coordinates $\{x,y,z\}$ on those time slices. 
At each step of the
evolution, compute the 3-metric $\gamma_{ij}$ and extrinsic curvature $K_{ij}$
of the slice $t = \hbox{constant}$ from the test-bed 4-metric $\{A, B^i,
\Gamma_{ij}\}$.  An acceptable lapse-shift prescription will keep $\alpha$,
$\beta$, $\gamma_{ij}$, and $K_{ij}$ all 
evolving on the slow timescale $\tau_*$;
and most likely will achieve this by maintaining $t\simeq T$,
$\alpha \simeq A$, $\beta^i \simeq B^i$, and $\gamma_{ij} \simeq \Gamma_{ij}$,
aside from fractional differences of order $1/(\Omega\tau_*)$.

\section*{Acknowledgments}

I thank Patrick Brady, Bernd Br\"ugman, and Jolien Creighton for helpful
conversations. This research was supported in part by NSF Grants
AST-9731698 and PHY-9424337, and NASA grant NAG5-6840.

\end{document}